\documentclass{elsart}
\usepackage{graphicx,amssymb,amsmath}
\begin{document}
\begin{frontmatter}

\title{Spin Bath Decoherence Mediated by Phonons}

\author{\"{O}zg\"{u}r Bozat},
\author{Zafer Gedik\corauthref{cor}}%
\corauth[cor]{Corresponding author. Tel.: +90 216 4839610; fax: +90
216 4839550.} \ead{gedik@sabanciuniv.edu}

\address{ Faculty of Engineering and Natural Sciences, Sabanci
University, Tuzla, Istanbul 34956, Turkey}

\begin{abstract}

We introduce an exactly solvable model to study decoherence of a
central spin interacting with a spin bath where the coupling is
mediated by phonons which we assume to be in a coherent state or
thermal distribution. For the coherent state case, we find that the
decoherence factor decays in a Gaussian fashion and it becomes
independent of the phonon frequencies at short times. If the phonon
energies are much larger than spin-phonon coupling or bath spins are
fully polarized, decoherence time becomes independent of the initial
phonon state. For the thermal state case, phonons play more
important role in decoherence with increasing temperature. We also
discuss possible effects of the temperature on spin bath
contribution to decoherence.

\end{abstract}

\begin{keyword}
A. Nanostructures \sep D. Spin dynamics \PACS 03.65.Yz \sep 03.67.-a
\sep 75.10.Jm \sep 75.75.+a \sep 76.20.+q
\end{keyword}
\end{frontmatter}

Decoherence is the key concept for understanding the emergence of
classical states out of a quantum system \cite{zurek2003}. This
phenomenon is also the main challenge in quantum information
processing~\cite{nielsen}. Coupling of central two level system
(qubit) to environment, leads to loss of phase relation between the
states of the qubit. Therefore, the superposition of the qubit
states evolves into a statistical mixture of the states, so called
pointer states. These states are determined by the form of the
system-environment interaction. If the qubit starts from a pure
state where it is decoupled from the environment, in time, the qubit
and the environment become quantum mechanically correlated. As the
qubit gets entangled with the environment, it can no longer be
described by a pure state. Although decoherence concept seems to
solve most of the puzzle of the emergence of classicality, there are
still open questions like "How does the information flow from system
to environment?". Understanding this information transfer is
believed to be crucial for explaining the objectivity of the
classical world \cite{zurek2006, zurek2005}.

Being the elementary quantum information units, qubits are one of
the most extensively studied open quantum systems. Especially solid
state qubits (quantum dots, SQUIDs, magnetic molecules, etc.) have
attracted a great interest due to their scalability which is an
indispensable criterion for realistic quantum information
processing. However, most important drawback of these systems is
their relatively strong couplings to the environment. Understanding
the mechanism of these interactions is crucial for the
implementation of error correction techniques \cite{shor95} and/or
error avoiding strategies \cite{lidar98}. Starting with the
pioneering works of Caldeira and Leggett \cite{Caldeira81,
Caldeira83},  the crucial effects of environment on the dynamics of
the central system has been studied with different models. Among
them spin-boson model has attracted much attention \cite{leggett87,
weiss99}. Now, it is a well understood environmental model. However,
this model is inadequate in most situations where localized
environmental modes act as a main source of decoherence
\cite{prokofev96, prokofev00}. In these cases spin bath models are
used to describe the environment. In spite of numerous theoretical
works, including both analytical approaches \cite{khaet02, merk02,
coish04, deng06, camalet07} and numerical simulations \cite{schli02,
tessi03, dobro03, zhang07}, spin bath decoherence is still a hot
subject. This is due to the rich dynamics of spin models with
different intra-bath couplings. Quantum dots are extensively studied
systems, both theoretically and experimentally, where the hyperfine
interaction with nuclear spins is dominating mechanism of
decoherence \cite{qdot}.

Generally, bosonic and fermionic modes are considered to be
effective at different time scales and they are coupled to the qubit
independently. However, this is not always the case. For instance,
recent experiments on particular single molecular magnets show that
these two mechanisms cooperate together \cite{evan04, more04}. It
has been proposed that the Waller mechanism, modulation of the
dipolar fields by atomic vibrations, can play an important role
\cite{waller}. Phonon assisted hyperfine interaction in quantum dots
is another example \cite{Abalmassov}. Deviations of the nucleus
positions due to lattice vibrations modify the hyperfine coupling
with electron spin. A final example can be given from optical
lattices where coupling strengths between spins trapped deep inside
a confining potential change with lattice oscillations. Starting
with the theoretical study of Jaksch \emph{et al.} \cite{jaksch},
ultra-cold atomic gases in optical lattices have attracted great
attention. Possibility of controlling the interactions among trapped
particles is most advantageous property of these systems. This
peculiarity enables to mimic various spin models such as Ising, XY,
Heisenberg and so on (see for review \cite{Lewenstein}). These
developments lead to various applications in quantum information
processing \cite{cirac} and study of spin bath decoherence in a
controlled way \cite{ross07, duan03, jan03}.

Inspired by these observations, we introduce a pure dephasing model
where the interaction of the central two-level system with
environmental spins is mediated by phonons. Study of pure dephasing
model is motivated by two observations. Firstly, dissipative
processes where the energy exchange occurs between subsystems have
typically longer time scales than pure dephasing processes
\cite{abragam61}. Secondly, exact solubility of the model gives a
more clear understanding of the decoherence process. We neglect the
self-Hamiltonians of the central spin and the spin bath. In
particular, we don't consider any interaction among the bath spins.
We assume that low energy physics of the system is governed by the
effective Hamiltonian
\begin{equation}
H=c_{z}\sum_{k=1}^{N}\left[ \omega _{0k}+\omega _{k}\left(
p_{k}^{\dag }+p_{k}\right) \right] s_{kz}+\sum_{k=1}^{N}\Omega
_{k}p_{k}^{\dag }p_{k} \label{ham1}
\end{equation}
where $c_{z}$ and $s_{kz}$ are z-components of the Pauli spin
operators for central two-level system and $k^{th}$ spin of the
bath, respectively. $N$ is the total number of environmental spins.
We are using units such that the Planck and the Boltzmann constants
are unity. $p_{k}^{\dag }$ and $p_{k}$ are the boson creation and
annihilation operators with commutation relation $
[p_{k},p_{k^{\prime }}^{\dag }]=\delta _{k,k^{\prime }}$. Energy
eigenvalues of the phonon bath are $\Omega _{k}$, and the coupling
strengths are $\omega _{0k}$ and $\omega _{k}$. Our model is similar
to the one proposed by Zurek where the central two level system is
directly coupled to spin bath \cite{zurek1982}. In our model this
coupling occurs with the help of oscillatory modes. When $\omega
_{k}$ and $\Omega _{k}$ vanish our model reduces to Zurek's. The
model Hamiltonian describes a system where the interaction between
the central two-level system and the bath spins is distance
dependent and this distance is modified by some vibrational modes.

First, we solve the case where the qubit is surrounded by spins
almost localized at different positions, for example at lattice
points of a solid. The interaction strengths between the system and
a bath spins change with the distance between them. Considering the
displacement of these atomic positions as macroscopic vibrations, we
model them by coherent states which are the most classical states of
phonons. An atom, confined in a harmonic potential, satisfies the
minimum position-momentum uncertainty when it is in a coherent state
which is nothing but a Gaussian wave function displaced from the
origin. Furthermore, it oscillates while preserving its shape, i.e.
it remains as a coherent state. We assume that initially the system
and the environment are uncorrelated so that the initial wave
function can be written as a product state,
\begin{equation}
|\Psi (0)\rangle =\left( c_{\uparrow }|\uparrow \rangle
+c_{\downarrow }|\downarrow \rangle \right)
\bigotimes_{k=1}^{N}\left( \alpha _{k}|\uparrow _{k}\rangle +\beta
_{k}|\downarrow _{k}\rangle \right) |\lambda _{k}\rangle \label{psi}
\end{equation}
where $|\uparrow \rangle (|\uparrow _{k}\rangle )$ and $|\downarrow
\rangle (|\downarrow _{k}\rangle )$ are normalized eigenstates of
$c_{z}(s_{kz})$ with eigenvalues +1 and -1, respectively. Expansion
coefficients satisfy $ |c_{\uparrow }|^{2}+|c_{\downarrow
}|^{2}=|\alpha _{k}|^{2}+|\beta _{k}|^{2}=1$ so that $|\Psi
(0)\rangle $ is normalized. $|\lambda _{k}\rangle $ is the coherent
state corresponding to the annihilation operator $p_{k}$ with
eigenvalue $\lambda _{k}$ so that $p_{k}|\lambda _{k}\rangle
=\lambda _{k}|\lambda _{k}\rangle $. With the help of the harmonic
displacement operators $D(\alpha )=e^{\alpha p^{\dagger }-\alpha
^{\ast }p}$, Hamiltonian can be diagonalized easily. Applying the
propagator $e^{-itH}$, we can calculate the time evolution of the
wave function which can be written as $|\Psi (t)\rangle =c_{\uparrow
}|\uparrow \rangle |B_{+}(t)\rangle +c_{\downarrow }|\downarrow
\rangle |B_{-}(t)\rangle $ where
\begin{equation}
|B_{\pm }(t)\rangle =\bigotimes_{k=1}^{N}\left( \alpha
_{k}A_{k}^{\pm }|\uparrow _{k}\rangle |u_{k}^{\pm }\rangle +\beta
_{k}A_{k}^{\mp }|\downarrow _{k}\rangle |u_{k}^{\mp }\rangle
\right). \label{psienv}
\end{equation}
Here $|u_{k}^{\pm }\rangle $ are the coherent states with
eigenvalues
\begin{equation}
u_{k}^{\pm }=(\lambda _{k}\pm \frac{\omega _{k}}{\Omega
_{k}})e^{-it\Omega _{k}}\mp \frac{\omega _{k}}{\Omega _{k}},
\end{equation}and
\begin{equation}
A_{k}^{\pm }=e^{i\frac{\omega _{k}^{2}}{ \Omega _{k}}\left(
t-\frac{\sin (\Omega _{k}t)}{\Omega _{k}}\right) }e^{\mp it\omega
_{0k}}e^{\mp i \frac{\omega _{k}}{\Omega _{k}}\left(
\text{Re}[\lambda _{k}]\sin (\Omega _{k}t)+\text{Im}[\lambda
_{k}](1-\cos (\Omega _{k}t))\right) }.
\end{equation}
Total density matrix is given by $\rho =|\Psi (t)\rangle \langle
\Psi (t)|$. Reduced density matrix of the central system $\rho _{c}$
is obtained by tracing over all the environmental degrees of freedom
as $\rho _{c}=Tr_{bath}\rho $. In $c_{z}$-basis, the reduced density
matrix is given by
\begin{equation}
\rho _{c}=\begin{pmatrix}
  |c_{\uparrow}|^{2} & c_{\uparrow}c_{\downarrow}^{\ast}r \\
  c_{\uparrow}^{\ast}c_{\downarrow} r^ {\ast} & |c_{\downarrow}|^{2} \\
\end{pmatrix}
\end{equation}
Magnitude of the off-diagonal matrix element is determined by the
decoherence factor
\begin{equation}
r(t)=\prod_{k=1}^{N}(|\alpha _{k}|^{2}A_{k}^{-^{\ast
}}A_{k}^{+}\langle u_{k}^{-}|u_{k}^{+}\rangle +|\beta
_{k}|^{2}A_{k}^{+^{\ast }}A_{k}^{-}\langle
u_{k}^{+}|u_{k}^{\_}\rangle )  \label{rho1}
\end{equation}
which can be written more explicitly as
\begin{eqnarray}\label{r0}
r(t) =\prod_{k=1}^{N}e^{-4\frac{\omega _{k}^{2}}{\Omega
_{k}^{2}}(1-\cos (\Omega _{k}t))}&&~(|\alpha _{k}|^{2}e^{-i2\omega
_{0k}t-i4\frac{\omega _{k} }{\Omega _{k}}(\text{Re}[\lambda
_{k}]\sin (\Omega
_{k}t)+\text{Im}[\lambda_{k}] (1-\cos (\Omega _{k}t)))}\nonumber \\
&&+|\beta _{k}|^{2}e^{i2\omega _{0k}t+i4\frac{\omega _{k}}{\Omega
_{k}}(\text{Re}[\lambda _{k}]\sin (\Omega _{k}t)+\text{Im}[\lambda
_{k}](1-\cos (\Omega _{k}t)))}). \nonumber \\
&&\label{rho2}
\end{eqnarray}

At $t=0$, $r=1$ and as $t$ increases, in general, it decays to zero
which means that interference of the states $|\uparrow \rangle $ and
$|\downarrow \rangle $ is totally suppressed. At short enough times
we can expand the trigonometric functions by treating $\Omega
_{k}t$'s as small parameters to obtain
\begin{equation}
r(t)\approx \prod_{k=1}^{N}e^{-2\omega _{k}^{2}t^{2}}~(|\alpha
_{k}|^{2}e^{-it(4\omega _{k}\text{Re}[\lambda _{k}]+2\omega
_{0k})}+|\beta _{k}|^{2}e^{it(4\omega _{k}\text{Re}[\lambda
_{k}]+2\omega _{0k})}). \label{rho3}
\end{equation}
If either the coupling strengths $\omega _{k}$'s and $\omega
_{0k}$'s or coherent state eigenvalues $\lambda _{k}$'s are random
enough, the second factor in the product leads to further
suppression of the coherence factor so that $r$ decays in Gaussian
form for large $N$ \cite{cucc05},
\begin{equation}
|r(t)|\approx e^{-t^{2}\sum_{k}(8|\alpha _{k}|^{2}|\beta
_{k}|^{2}(2\omega _{k}\text{Re}[\lambda _{k}]+\omega
_{0k})^{2}+2\omega _{k}^{2})}. \label{rho4}
\end{equation}
Therefore phonon energies do not play any role for short time
decoherence of the central system. The decoherence time is
determined by the coupling constants and the initial configurations
of the phonons and spin bath states. It is interesting to note that
even if all the bath spins are polarized in one direction, i.e.
$|\alpha _{k}|^{2}=1$, system still losses its coherence. This
behavior is a result of presence of the phonons in the environment.
It is also interesting that phonon state eigenvalues ($\lambda
_{k}$'s) do not affect the decoherence time in this case. It is
obvious that in the limit of $\Omega _{k}\rightarrow 0$ and $\omega
_{k}\rightarrow 0$, our Hamiltonian is reduced to Zurek's model
where decoherence is due to direct spin-spin interactions only
without phonon contribution. In this case initial configuration of
the spin bath becomes crucial.

Another interesting case is the $\Omega _{k}/\omega_k\rightarrow
\infty $ limit where the decoherence factor
$r(t)=\prod_{k=1}^{N}(|\alpha _{k}|^{2}e^{-i2\omega _{0k}t}+|\beta
_{k}|^{2}e^{i2\omega _{0k}t})$. Therefore, $r(t)$ depends on the
initial configuration of bath spins only and it becomes independent
of the initial phonon state eigenvalues. Since the separation of
energy levels of phonons becomes very high, phonon states do not
change in time and remain uncorrelated to system and bath spins.

Now, we analyze the case where phonons are in a thermal equilibrium
rather than a coherent state. Such a situation can physically be
realized when the atoms carrying bath spins are brought in contact
with a heath bath to thermalize before $t=0$. For thermal states
phonon density matrix is given by
\begin{equation}
\rho_{p}(0)=\bigotimes_{k}^{N}(1-e^{-\frac{\Omega_{k}}{T}})\sum_{n_{k}=0}^{\infty}e^{-\frac
{\Omega_{k}n_{k}}{T}}|n_{k}\rangle\langle n_{k}|.
\end{equation}
We assume that the bath spins are in a separable state at $t=0$ as
before. Since in the Hamiltonian there are no intra-bath terms for
the spins, heath bath thermalizing the phonons will simply randomize
the initial spin directions. As we shall discuss below, if bath
spins have individual energy levels for up and down configurations,
heath bath will determine the initial occupation numbers for the two
possible states in accordance with the Gibbs factors. Time evolution
of the total density matrix is given by
$\rho(t)=e^{-iHt}\rho(0)e^{iHt}$. Using the over-completeness
relation
\begin{equation} 1=\frac{1}{\pi}\int
d^{2}\lambda|\lambda\rangle\langle\lambda|,
\end{equation}
and the number state representation of coherent states
\begin{equation}\langle
n|\lambda\rangle=e^{-|\lambda|^{2}/2}\frac{\lambda^{n}}{\sqrt{n!}},
\end{equation}
it is straight forward to calculate the reduced density matrix of
the central system. In this case decoherence factor becomes
\begin{equation}
\label{r2}
r(t)=\prod_{k=1}^{N}e^{-4\frac{\omega_{k}^{2}}{\Omega_{k}^{2}}(1-\cos
(\Omega_{k}t))\coth(\frac{\Omega_{k}}{T})}
(|\alpha_{k}|^{2}e^{-i2\omega_{0k}
t}+|\beta_{k}|^{2}e^{i2\omega_{0k}t}).
\end{equation}
We first note that for $\Omega_k/T\rightarrow \infty$, Eq.
(\ref{r2}) and Eq. (\ref{r0}) become identical provided that
$\lambda_k=0$. This is a consistency check for two phonon states,
coherent states and thermal states, that we have discussed because
at low temperatures thermal state approaches the ground state of the
harmonic oscillators which are nothing but the coherent states with
vanishing eigenvalues.

According to Eq. (\ref{r2}), decoherence factor has two
contributions, coming from phonons and spins. The two mechanisms act
simultaneously in decoherence of the central spin. Depending upon
the interaction strengths, one of them can become the dominant
mechanism. At very low temperatures, where the hyperbolic cotangent
term is approximately unity, the first term becomes independent of
$\Omega_k$ values provided that $t$ is small enough. For large
temperatures, decoherence factor becomes an exponentially decaying
function of $T$. It is possible to generalize the model Hamiltonian
by adding a $s_{kz}$-dependent intra-bath term for individual spins.
In this case the heath bath will not only thermalize the phonons but
also it will determine the $|\alpha_{k}|^2/|\beta_{k}|^2$ ratio. For
example, at very large temperatures the ratio will tend to unity and
hence the spin bath will have a more important contribution to
decoherence in comparison to lower temperatures.

In conclusion, we examined a spin decoherence model where the
interaction with the bath spins are modified by phonons. Coherent
states of phonons correspond to almost localized bath spins. In this
case we find that initial decoherence rate does not depend on the
phonon energies. Furthermore, for polarized bath spins it becomes
independent of the initial phonon states. Thermal phonon
distribution is the other case where we found an explicit solution
of the model Hamiltonian. At high temperatures phonons play a more
important role in the dephasing process.

\begin{ack}
This work has been supported by the Scientific and Technological
Research Council of Turkey (TUBITAK) under grant 107T530.
\end{ack}

\end{document}